# SPATIAL MODELING OF MENTAL HEALTH ON OUTPATIENT MORBIDITY IN KENYA

Ndegwa, RW[1], Mwalili, S [1], Wamwea, C[1]

[1]Jomo Kenyatta University of Agriculture and Technology (JKUAT), Juja, Nairobi, Kenya; Department of Statistics and Actuarial Science, P.O Box Thika
Corresponding Author; Ndegwa, RW
Email; Wambundegwa99@gmail.com
Mobile: +254704905333

**Abstract**

A mental health disorder is a clinically significant impairment in a person's intellect, emotional control, or behavior.Mental disorders and outpatient morbidity are a challenge to public health in Kenya.The spatial distribution and study of factors associated with these conditions remain limited. The study aimed to conduct spatial modeling of mental health on outpatient mobility in Kenya. This project used spatial modeling to explore the relationship between infectious diseases and mental disorders. The results showed that mental health issues were not distributed uniformly, with higher frequency found in Western and Nairobi regions. Possible connections between HIV, TB, STIs and mental health have been suggested by the substantial correlation that has been found between infectious diseases and mental health issues. The spatial model that was created demonstrated excellent validity and accuracy, providing policymakers with a useful tool to better allocate resources and enhance mental health treatments, especially in high-risk locations. In conclusion, the research improved knowledge of the spatial patterns of mental health disorders and guides intervention tactics and healthcare policies in Kenya and other comparable settings. There is a need for development and implementation of geographically tailored mental health intervention programs should be done in accordance with the high-prevalence areas.

**Introduction**

Psychiatric diseases, often known as mental illnesses or mental health disorders, have been a major global health concern for both people and public health systems. These illnesses interfered with people's ability to think clearly and make decisions by warping their views and mental processes. Mental Health Evidence and Research Team (2005) I have highlighted the crucial role that mental health plays in overall well-being. They have highlighted the notion that "no health without mental health" (Prince et al., 2007).

In Kenya, Public health systems are heavily burdened by mental health problems, but there is a key knowledge gap about the regional distribution of these diseases and their effect on outpatient morbidity (Meyer & Ndetei, 2016). The creation of workable healthcare policies and efficient resource management techniques is impeded by this knowledge gap. Addressing these problems and advancing equitable mental health treatment across various regions required an understanding of the spatial patterns of mental health and their implications for outpatient morbidity (Tadmon & Bearman, 2023).

Untangling the link between mental health problems and outpatient morbidity, which enables a thorough knowledge of the influence of mental health on overall healthcare consumption, is another key component .World Health Organization. (2001) It's crucial to understand whether particular mental health issues are associated with particular kinds of outpatient morbidity. Janse Van Rensburg et al. (2020) confirmed that such information would help to promote a holistic approach to healthcare delivery by shedding light on the complex interactions between physical and mental health.

Spatial modeling methods to examine outpatient morbidity and mental health in Kenya may provide insight into potential risk factors or drivers influencing the patterns that have been noticed (Fisher & Langford, 1995). We can discover the complex influences on mental health outcomes as discussed by Porter (2019)by examining healthcare-related factors across various geographic areas.

The study aimed at spatial modelling to evaluate the geographic distribution of mental health issues, determining the connections between these disorders and HIV, STIs, and Tuberculosis. In addition, assess the precision and soundness of the spatial model and to educate targeted therapists and policy decisions, as well as offer insights into the regional dynamics of mental health.

**Methodology**
**Ethical Considerations**
Before start of the study, the Jomo Kenyatta University of Agriculture and Technology (JKUAT) Institutional Scientific and Ethics Review Committee (ISEREC) granted ethical clearance, guaranteeing that it complied with the rules for research with human beings.

**To determine the spatial distribution of mental health in Kenya**

The Intrinsic Conditional Autoregressive (ICAR) model was employed to investigate spatial dependence in the data. Key components included:

- A collection of non-overlapping areal units represented as $S = \{S_1,...,S_K\}$, each with known offsets $O = \{O_1,...,O_K\}$ and corresponding responses $Y = \{Y_1,...,Y_K\}$.
- A matrix of covariates $X = \{x_1,...,x_K\}$ and a spatial structure component $\psi = \{\psi_1,...,\psi_K\}$, representing residual spatial autocorrelation. • The relationship between covariates and responses modeled using a hierarchical Bayesian setting:

$Y_k|\mu_k \sim f(y_k|\mu_k, v^2)$, where $\beta \sim N(\mu_\beta, \Sigma_\beta)$ and $\mu_k = x_k^T \beta + O_k + \psi_k$.

- Modeling of the spatial structure component $\psi$ using a weight or adjacency function.

**ICAR Models for Areal Data**

**Framework for CAR Model** Set of $k = 1,...,K$ non-overlapping areal units $S = S_1,...,S_K$ linked to corresponding responses $Y = (Y_1,...,Y_K)$ and known

(1)

Spatial variation in the response is modeled by a matrix of covariates $X = (x_1,...,x_K)$ and a spatial structure component $\psi = (\psi_1,...,\psi_K)$, where the latter is included to model any remaining spatial autocorrelation after accounting for covariate effects. The vector of covariates for areal unit $S_k$ is denoted by $x_k = (1, x_{k1},...,x_{kp})$. Using the Hierarchical Bayesian Setting:

$$Y_k|\mu_k \sim f(y_k|\mu_k, v^2), \quad k = 1,...,K \qquad (2)$$

$$g(\mu_k) = x_k^T \beta + O_k + \psi_k \qquad (3)$$

where $\beta \sim N(\mu_\beta, \Sigma_\beta)$. Within $\psi$, a weight or adjacency matrix controls spatial autocorrelation.

**To evaluate the relationship between mental health conditions and HIV and TB**
**Structure of the Conditional Autoregressive (CAR) Model**
The study focused on disease risk assessments using the count ($y$) of infection cases with the following model structure:

$y_i \sim \text{Pois}(\lambda_i E_i)$, $\eta_i = \log(E_i) + \log(\lambda_i)$,

$$\log(\lambda_i) = \beta_0 + \sum_{j=1}^{p} \beta_j x_{ij} + s_i, \qquad i = 1,2,\ldots,N,$$

where $E_i$ is the offset of polygon $i$, $\lambda_i$ is the relative risk of diseases, $\beta_0$ is the intercept, $\beta_j$ is the coefficient of covariate $x_{ij}$ in polygon $i$, and $s_i$ represents the spatially structured variate of polygon $i$ following an ICAR distribution.

The ICAR model captures spatial dependency among neighboring polygons. The distribution of $s_i$ follows an ICAR distribution, mathematically represented as:

$$s_i|s_{-i} \sim \mathcal{N}\left(\sum_{j=1}^{n} w_{ij} s_j, \tau_i^2\right)$$

where $s_{-i}$ represents the set of spatially structured variates of all polygons except polygon $i$, $w_{ij}$ represents the spatial weights specifying the connectivity between polygon $i$ and polygon $j$, and $\tau_i^2$ is the precision parameter of the ICAR model for polygon $i$.

The ICAR model assumes that the spatially structured variate of each polygon is influenced by the spatially structured variates of its neighboring polygons. The spatial weights $w_{ij}$ are typically defined based on adjacency or contiguity between polygons.

By incorporating the ICAR model structure into the analysis of disease risk assessments, this study aims to account for spatial autocorrelation in count data and capture underlying spatial patterns in disease occurrence. The ICAR model allows estimation of the relative risk of diseases while considering spatial dependencies among neighboring polygons.

**Disease mental risk Mapping using Conditional Autoregressive Besag-York-Mollié( CAR BYM)**

For disease risk mapping, the Conditional Autoregressive-BYM (CAR-BYM) model was utilized. This model comprised a regular random effect component for spatial rarefaction and an ICAR component for spatial autocorrelation.

The model equation was:

$$\eta_i = \mu + \mathbf{x}_i'\boldsymbol{\beta} + \phi_i + \theta_i \tag{4}$$

where $\mu$ represented the average risk level, $\mathbf{x}_i$ denoted the observational vector of independent variables for unit $i$, $\beta$ was the parameter vector, $\phi_i$ represented the ICAR component, and $\theta_i$ represented the random effect of non-spatial heterogeneity.

**To evaluate accuracy and validity of the spatial model used**

**Bayesian Inference with MCMC**

Markov Chain Monte Carlo (MCMC) techniques were employed for Bayesian parameter estimation. Samples from the joint posterior distribution were generated iteratively using the Gibbs sampler method.

The algorithm iterated as follows for each parameter $\theta$:

1. Sample $\theta_t^{(1)}$ from $p(\theta^{(1)}|\theta_{t-1}^{(2)},...,\theta_{t-1}^{(k)},\mathbf{y})$.

2. Sample $\theta_t^{(2)}$ from $p(\theta^{(2)}|\theta_t^{(1)},\theta_{t-1}^{(3)},...,\theta_{t-1}^{(k)},\mathbf{y})$.

3. ...

4. Sample $\theta_t^{(k)}$ from $p(\theta^{(k)}|\theta_t^{(1)},...,\theta_{t-1}^{(k-1)},\mathbf{y})$.

Equation 4 represents the model equation for the CAR-BYM model.

Convergence was assessed, and parameter estimates were obtained by calculating the posterior means.

Evaluation of the Model:

**Information Criterion for Deviance (DIC)**:

Model fit was measured using the Information Criterion for Deviance (DIC0, which was also utilized for model selection. The formula for computing it was:

$$DIC = D(\theta) + P_b,$$

in which $P_b$ denoted a Bayesian measure of model complexity and $D(\theta)$ represented the posterior mean of the deviance. A lower DIC value suggested that the model suited the data better.

**Model Implementation**:

Using ICAR models and Bayesian inference techniques, this research offered a comprehensive framework for investigating spatial dependency in mental health outcomes and evaluating the influence of factors connected to infectious diseases.

### 3. Data Analysis

In this section, the research conducted a spatial mapping of mental illnesses in Kenya and explored the relationship between HIV, STIs, and TB with respect to mental disorders.

**Data Collection** Data on mental health was sourced from medical institution websites, data collection firms, publications, and data documents. Secondary sources included the Institute of Health Metrics and Evaluation, the Kenya National Bureau of Statistics, and the Mathari National Teaching and Referral Hospital.

**Software for Data Analysis** The software of choice was R-Studio and winBUGS as they contain in built tested out functions required to carry out Data Preprocessing, Exploratory Data Analysis, Parameter Estimation and Inference by utilizing Markov Chain Monte Carlo.

## Results

**Spatial distribution of mental health conditions among individuals seeking outpatient care in different regions of Kenya**

**Mental Illness Incidence**

The study results map showed that the prevalence of mental illnesses is highest in the urban areas of Kenya. The counties with the highest (37.8-141) prevalence of mental illnesses are Nairobi, Mombasa, Kiambu, Kisumu, and Nakuru. Conversely, the counties with the lowest (1-10.2) prevalence of mental illnesses are Mandera, Wajir, Garissa, Tana River, and Turkana (Fig. 1).

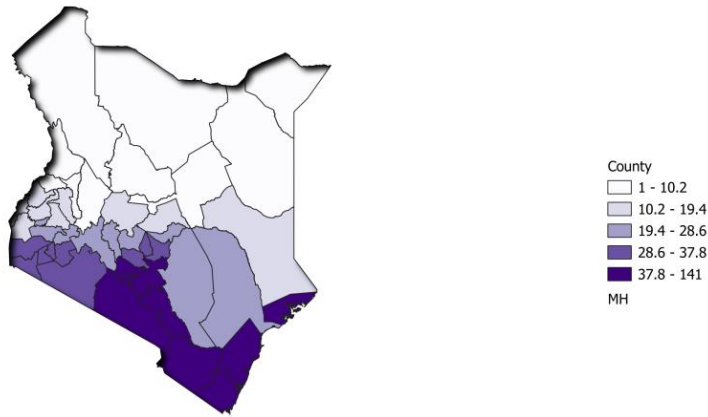

Figure 1. Mental illness incidence
Source: Ndegwa, 2023

**STI Incidence**

The counties with the (37.8-141) highest prevalence of STIs are located in the western and southwestern parts, including Siaya, Kisumu, Migori, Homa Bay, Busia, Bungoma, Bomet, Nandi, and Trans Nzoia. In addition, there were pockets of high STI prevalence (37.8-141) in other parts such as Nairobi and Kilifi County (Fig.2).

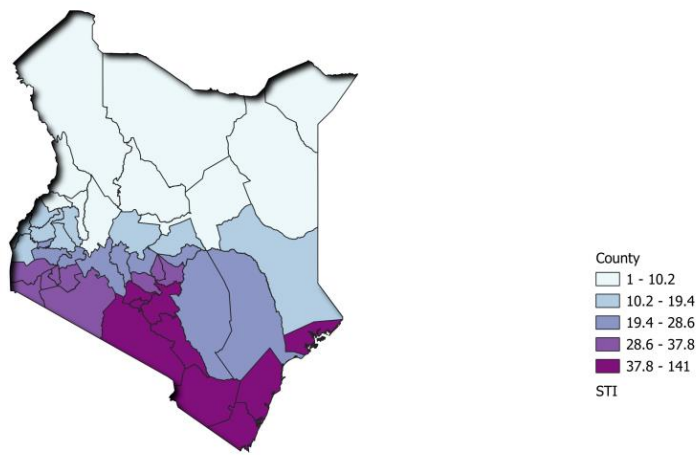

Figure 2: STI incidence

Source: Ndegwa, R.W, 2023

**HIV Incidence**

Counties with the highest (6.9-20.6) rates of HIV prevalence include Siaya, Kisumu, Migori, Homa Bay, Busia, Bungoma, Nyamira, Kisii, Kericho, Bomet, Nandi, and Trans Nzoia. Other parts such as Nairobi County and Kilifi County also exhibit significant (5.5-6.9) HIV prevalence. Conversely, the lowest HIV prevalence rates (0-1.5) are found in Marsabit, Wajir, Garissa, Mandera, Isiolo, Tana River, Turkana, Lamu, Samburu, West Pokot, Elgeyo Marakwet, and Baringo (Fig.3).

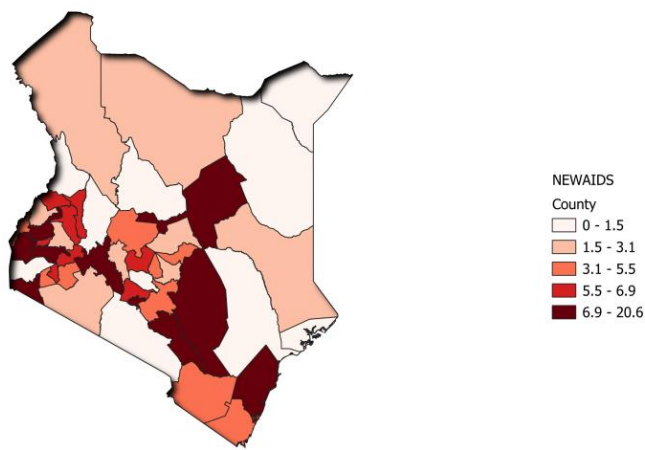

**Figure 3**: HIV incidence
Source: Ndegwa, 2023

**Tuberculosis Incidence**
Darker colouring on the map of Kenya denoted higher tuberculosis incidence(0.00098-0.00139). The western and Central regions had greater rates of tuberculosis than the Coastal and North Eastern regions (Fig. 4).

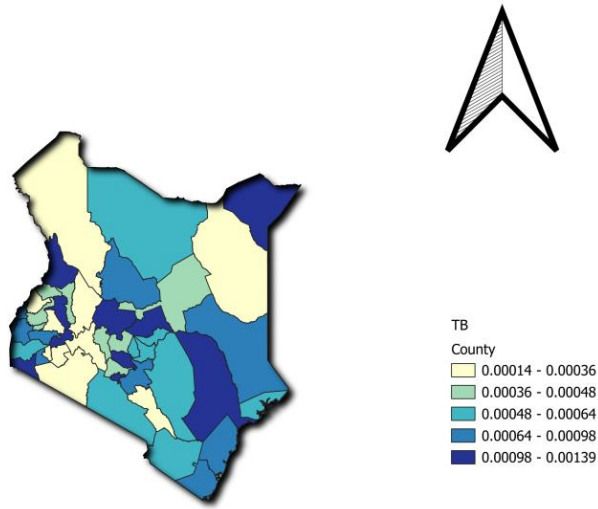

**Figure 4:** Tuberculosis incidence
Source: Ndegwa, 2023

**Evaluating the relationship between mental health conditions and HIV and TB at the regional level.**

**Non-spatial analysis**

1. **STI (Infections Transmitted by Sexual Intercourse - $\alpha_1[1]$):**

    - The effect of STI on mental illness ($\alpha_1[1]$) has a posterior mean estimate of around 0.005517.

    - The 95% credible interval (CI) for $\alpha_1[1]$ ranges from 0.004642 to 0.005518.

**Inference**: STIs and mental illness are positively and statistically significantly correlated (0.005517).

**Table 1:** Non-spatial analysis on Relationship between Mental Disorders and HIV, TB, and STIs

| Node | Mean | SD | MC Error | 2.5% | Median | 97.5% | Start | Sample |
|---|---|---|---|---|---|---|---|---|
| $\alpha_0$ | 0.05076 | 0.03156 | $6.435 \times 10^{-4}$ | -0.01194 | 0.05104 | 0.1111 | 1 | 20000 |
| $\alpha_1[1]$ (STI) | 0.005517 | $4.485 \times 10^{-4}$ | $8.865 \times 10^{-6}$ | 0.004642 | 0.005518 | 0.00638 | 1 | 20000 |
| $\alpha_1[2]$ (TB) | $6.916 \times 10^{-4}$ | $1.998 \times 10^{-4}$ | $2.558 \times 10^{-6}$ | $2.978 \times 10^{-4}$ | $6.934 \times 10^{-4}$ | 0.001082 | 1 | 20000 |
| $\alpha_1[3]$ (AIDS) | -0.02062 | 0.002468 | $0.103 \times 10^{-5}$ | -0.0255 | -0.02058 | -0.01587 | 1 | 20000 |

2. $\alpha_1[2]$ **(Tuberculosis)**:

   - The posterior mean estimate for the impact of tuberculosis on mental illness ($\alpha_1[2]$) is approximately $6.916 \times 10^{-4}$.

   - The 95% credible interval (CI) for $\alpha_1[2]$ spans roughly from $2.978 \times 10^{-4}$ to $6.934 \times 10^{-4}$.

**Inference**: There exists a statistically significant and positive association between mental illness and tuberculosis. (data) Specifically, there is a slight rise in mental illness for every unit increase in tuberculosis within the credible interval.

3. $\alpha_1[3]$ **(AIDS)**:

   - The effect of AIDS on mental illness ($\alpha_1[3]$) has a posterior mean estimate of about $-0.02062$.

   - The 95% credible interval (CI) for $\alpha_1[3]$ ranges from $-0.0255$ to $-0.01587$.

**Inference**: There is a statistically significant and adverse correlation between mental illness and HIV. Specifically, there is a correlation between a one-unit rise in HIV and a decline in mental illness that is contained within the credible interval (Table 1)

**Spatial Analysis**

Below is a table showing the relationship between Mental Disorders and HIV, TB, and STIs with the spatial aspect in consideration.

**Table 2**: Relationship between Mental Disorders and Infectious Diseases

| Node | Mean | SD | MC Error | 2.5% | Median | 97.5% | Start | Sample |
|---|---|---|---|---|---|---|---|---|
| $\alpha_0$ | -0.1441 | 0.06276 | 0.002417 | -0.269 | -0.143 | -0.02424 | 1 | 20000 |
| $\alpha_1[1]$ (STI) | 0.002576 | $6.686 \times 10^{-4}$ | $2.023 \times 10^{-5}$ | 0.001274 | 0.002576 | 0.003872 | 1 | 20000 |
| $\alpha_1[2]$ (TB) | $5.935 \times 10^{-4}$ | $2.483 \times 10^{-4}$ | $5.141 \times 10^{-6}$ | $1.037 \times 10^{-4}$ | $5.954 \times 10^{-4}$ | 0.001074 | 1 | 20000 |
| $\sigma$ | 1.462 | 0.1665 | 0.00168 | 1.175 | 1.448 | 1.83 | 1 | 20000 |
| $\tau$ | 0.4861 | 0.1086 | 0.001118 | 0.2987 | 0.4768 | 0.7241 | 1 | 20000 |
| $\alpha_1[3]$ (AIDS) | 0.002954 | 0.00654 | $2.265 \times 10^{-4}$ | -0.009481 | 0.002846 | 0.01576 | 1 | 20000 |

1. $\alpha_0$:

    - The posterior mean estimate for $\alpha_0$ is approximately -0.1441.
    - The standard deviation (SD) of $\alpha_0$ is approximately 0.06276.
    - The 95% credible interval for $\alpha_0$ ranges from approximately -0.269 to -0.02424.

**Inference**: In the context of this analysis, $\alpha_0$ represents the baseline effect on mental illness when all other covariates (STI, TB, AIDS) are zero. The posterior mean of approximately -0.1441 suggests a negative impact on mental illness, and the 95% credible interval indicates that this effect is statistically significant and falls between -0.269 and -0.02424.

2. $\alpha_1[1]$ (associated with STI):

    - The posterior mean estimate for $\alpha_1[1]$ is approximately 0.002576.
    - The standard deviation (SD) of $\alpha_1[1]$ is approximately $6.686 \times 10^{-4}$.
    - The 95% credible interval for $\alpha_1[1]$ ranges from approximately 0.001274 to 0.003872.

**Inference**: $\alpha_1[1]$ represents the effect of STI on mental illness. The posterior mean of approximately 0.002576 suggests a positive impact, and the 95% credible interval indicated that this effect was statistically significant and fell between 0.001274 to 0.003872.

3. $\alpha_1[2]$ (associated with TB):

    - The posterior mean estimate for $\alpha_1[2]$ is approximately $5.935 \times 10^{-4}$.
    - The standard deviation (SD) of $\alpha_1[2]$ is approximately $2.483 \times 10^{-4}$.
    - The 95% credible interval for $\alpha_1[2]$ ranges from approximately $1.037 \times 10^{-4}$ to 0.001074.

**Inference**: In the context of this analysis, $\alpha_1[2]$ represents the effect of TB on mental illness. The posterior mean of approximately $5.935 \times 10^{-4}$ suggests a positive impact, and the 95% credible interval indicated that this effect was statistically significant and fell between $1.037 \times 10^{-4}$ to 0.001074.

4. $\alpha_1[3]$ (associated with AIDS):

- The posterior mean estimate for $\alpha_1[3]$ is approximately 0.002954.
- The standard deviation (SD) of $\alpha_1[3]$ is approximately 0.00654.
- The 95% credible interval for $\alpha_1[3]$ ranges from approximately -0.009481 to 0.01576.

**Inference**: In the context of this analysis, $\alpha_1[3]$ represents the effect of AIDS on mental illness. The posterior mean of approximately 0.002954 suggests a positive impact, and the 95% credible interval indicates that this effect is statistically significant. However, the wide credible interval suggests some uncertainty about the magnitude of this effect.

5. $\sigma$ (Standard Deviation of Residuals):

- The posterior mean estimate for the standard deviation of residuals ($\sigma$) is approximately 1.462.

**Inference**: $\sigma$ represents the variability of the unexplained part of mental illness not captured by the covariates. A higher $\sigma$ indicates higher variability in mental illness not explained by the model.

6. $\tau$ (Precision or Inverse Variance of the Covariate Effects):

- The posterior mean estimate for $\tau$ is approximately 0.4861.

**Inference**: $\tau$ represents the precision of the covariate effects. A lower $\tau$ indicates more precise estimates of the covariate effects.

**Spatial-Temporal Analysis**

**Table 3**: Spatial-Temporal Analysis on the Relationship between Mental Disorders and Infectious Diseases

| Node | Mean | SD | MC Error | 2.5% | Median | 97.5% | Start | Sample |
|---|---|---|---|---|---|---|---|---|
| $\alpha_0$ | -0.2158 | 0.0754 | 0.003107 | -0.3732 | -0.2132 | -0.07168 | 1 | 20000 |
| $\alpha_1[1]$ (STI) | 0.002381 | $6.904 \times 10^{-4}$ | $2.226 \times 10^{-5}$ | 0.001046 | 0.002375 | 0.003754 | 1 | 20000 |
| $\alpha_1[2]$ (TB) | $5.014 \times 10^{-4}$ | $2.557 \times 10^{-4}$ | $5.463 \times 10^{-6}$ | $-3.15 \times 10^{-6}$ | $5.032 \times 10^{-4}$ | $9.979 \times 10^{-4}$ | 1 | 20000 |
| $\alpha_3$ | 0.02319 | 0.01347 | $4.002 \times 10^{-4}$ | -0.003645 | 0.02329 | 0.04951 | 1 | 20000 |
| $\sigma$ | 1.485 | 0.1695 | 0.001817 | 1.196 | 1.471 | 1.857 | 1 | 20000 |
| $\tau$ | 0.4709 | 0.1051 | 0.001133 | 0.2901 | 0.4623 | 0.6989 | 1 | 20000 |
| $\alpha_1[3]$ (AIDS) | 0.006874 | 0.006923 | $2.652 \times 10^{-4}$ | -0.006332 | 0.006733 | 0.02119 | 1 | 20000 |

1. $\alpha_0$:

- The posterior mean estimate for $\alpha_0$ is approximately -0.2158.
- The standard deviation (SD) of $\alpha_0$ is approximately 0.0754.

- The 95% credible interval for $\alpha_0$ ranges from approximately -0.3732 to -0.07168.

**Inference**: In the context of this analysis, $\alpha_0$ represents the baseline effect on mental illness when all other covariates (STI, TB, AIDS) are zero. The posterior mean of approximately -0.2158 suggests a negative impact on mental illness, and the 95% credible interval indicates that this effect is statistically significant and falls between -0.3732 and -0.07168.

(a) $\alpha_{1[1]}$ (associated with STI):

- The posterior mean estimate for $\alpha_{1[1]}$ is approximately 0.002381.
- The standard deviation (SD) of $\alpha_{1[1]}$ is approximately $6.904 \times 10^{-4}$.
- The 95% credible interval for $\alpha_{1[1]}$ ranges from approximately 0.001046 to 0.003754.

**Inference**: In the context of this analysis, $\alpha_{1[1]}$ represents the effect of STI on mental illness. The posterior mean of approximately 0.002381 suggests a positive impact, and the 95% credible interval indicates that this effect is statistically significant and falls between 0.001046 and 0.003754.

(b) $\alpha_{1[2]}$ (associated with TB):

- The posterior mean estimate for $\alpha_{1[2]}$ is approximately $5.014 \times 10^{-4}$.
- The standard deviation (SD) of $\alpha_{1[2]}$ is approximately $2.557 \times 10^{-4}$.
- The 95% credible interval for $\alpha_{1[2]}$ ranges from approximately $-3.15 \times 10^{-6}$ to $9.979 \times 10^{-4}$.

**Inference**: In the context of this analysis, $\alpha_{1[2]}$ represents the effect of TB on mental illness. The posterior mean of approximately $5.014 \times 10^{-4}$ suggests a positive impact, but it's important to note that the 95% credible interval includes values very close to zero. This indicates some uncertainty about the magnitude and statistical significance of this effect.

(c) $\alpha_{1[3]}$ (associated with AIDS):

- The posterior mean estimate for $\alpha_{1[3]}$ is approximately 0.006874.
- The standard deviation (SD) of $\alpha_{1[3]}$ is approximately 0.006923.
- The 95% credible interval for $\alpha_{1[3]}$ ranges from approximately -0.006332 to 0.02119.

**Inference**: In the context of your analysis, $\alpha_{1[3]}$ represents the effect of AIDS on mental illness. The posterior mean of approximately 0.006874 suggests a positive impact, and the 95% credible interval indicates that this effect is statistically significant. However, the wide credible interval suggests some uncertainty about the magnitude of this effect.

(d) $\alpha_3$:

- The posterior mean estimate for $\alpha_3$ is approximately 0.02319.

- The standard deviation (SD) of $\alpha_3$ is approximately 0.01347.
- The 95% credible interval for $\alpha_3$ ranges from approximately −0.003645 to 0.04951.

**Inference**: $\alpha_3$ represents the Temporal aspect of the model. The posterior mean estimate of approximately 0.02319 indicates a certain positive impact, and the credible interval suggests that this effect is statistically significant.

## 1.3     Evaluation of accuracy and validity of the spatial model

Model Comparison

**Variability ($\sigma$) and Precision ($\tau$)**

In the context of the analysis, $\sigma$ represents the standard deviation of the residuals in your model, indicating the amount of unexplained variability in mental illness not captured by the covariates. The values for $\sigma$ in each of the three analyses are as follows:

- Non-Spatial Analysis: $\sigma \approx 1.462$
- Spatial Analysis: $\sigma \approx 1.485$
- Spatial-Temporal Analysis: $\sigma \approx 1.485$

The relatively close values of $\sigma$ across all three analyses suggest that the unexplained variability in mental illness remains similar. A higher $\sigma$ value implies a greater degree of unexplained variability in mental illness not accounted for by the model.

**Precision ($\tau$)**

In the context of the project analysis, $\tau$ represents the precision or the inverse variance of the covariate effects. A lower $\tau$ value indicates more precise estimates of the covariate effects. The values for $\tau$ in each of the three analyses are as follows:

- Non-Spatial Analysis: $\tau \approx 0.4861$
- Spatial Analysis: $\tau \approx 0.4709$
- Spatial-Temporal Analysis: $\tau \approx 0.4709$

The consistent range of values for $\tau$ across all three analyses suggests that the precision of the estimates of covariate effects is similar in each model.

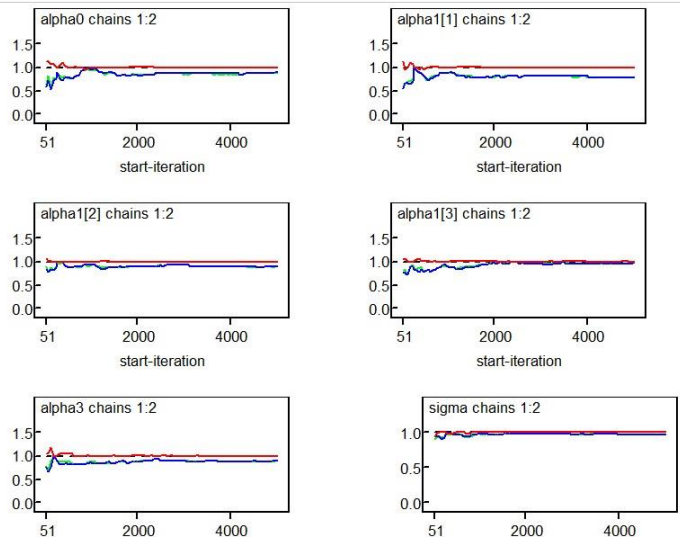

**Figure 5:** (9) Gelman Rubin Spatial-Temporal Trace

Following Convergence Criteria: Rhat Values: < 1.1, the model showed a High Degree of Convergence

**Model Comparison Based on DIC**

**Non-Spatial Model (GLM)**

- DIC = 0.0502
- Moderate model complexity
- Offers a decent match to the data

**Spatial Model**

- DIC = -0.1463
- Lower DIC compared to the non-spatial model
- Balances complexity and provides a better fit to the data

**Space-Time Model**

- DIC = -0.221
- Lowest DIC among the three models
- Offers the best balance between model complexity and data fitting

**4.Discussion (you just state the result in the order they follow each other, give reason for the finding and you agree or disagree with other scientists)**

The study results provide important new insights into a number of important aspects related to mental health issues among Kenyans seeking outpatient therapy. First off, various locations may require different levels of mental health care, as indicated by the non-uniform spatial distribution of mental health illnesses (Persad, 2020). The prevalence rates in urban areas were found to be greater than in rural areas, underscoring the significance of focused interventions in these areas. Moreover, the significant association shown between the prevalence of TB, HIV, and STIs and mental health disorders emphasizes the connection between the burden of infectious diseases and mental health problems. There are a number of possible reasons for the high prevalence of STIs in Nairobi and Kilifi. One reason may be that these counties have a high population density, which can lead to increased transmission of STIs. Another reason may be that these counties have high levels of poverty and unemployment, which can make it difficult for people to access healthcare and STI prevention services.There are considerable regional differences in the prevalence
of tuberculosis in Kenya, with the western and central regions of the nation having greater rates than the coastal and northeastern regions. These results emphasize the significance of comprehending the variables that influence TB disparities in Kenya and creating focused interventions to address these variables.High HIV prevalence in other parts of the country, such as Nairobi County and Homabay County was observed. This may be due to a number of factors, such as the high mobility of people in these areas, the high prevalence of commercial sex work, and the lack of access to healthcare and HIV prevention services.

Price et al. (1988) This result highlights the necessity of integrated healthcare strategies that address mental and physical health. The trustworthiness of the results is further strengthened by the validation of the geographic model employed in this investigation. The model offers useful information to policymakers regarding the allocation of resources and provision of services by precisely depicting the spatial patterns of mental health disorders and their correlation with infectious diseases.

**Conclusion**

The study concluded that The study found that among patients seeking outpatient therapy in various parts of Kenya, there is a non-uniform spatial distribution of mental health disorders . This non-uniform distribution suggests that the needs for mental health care may differ geographically. The study found a strong correlation at the regional level between the prevalence of HIV, STIs, and TB and mental health disorders.There may be a complex interaction between the burdens of infectious diseases and mental health issues because regions with greater rates of these infectious disease burdens also had higher rates of mental health disorders This study's spatial model is determined to be both valid and accurate, as it successfully pinpoints regions with higher rates of mental health issues. Healthcare planners can benefit greatly from the model by using it to better allocate resources and enhance access to mental health services in high-risk areas. The results highlight the value of focused therapies and integrated healthcare methods in addressing the intricate interactions between mental and physical health problems. The study recommends that the inclusion of mental health treatments into currently running HIV and TB healthcare initiatives should be encouraged to improve patient care and make it more comprehensive and efficient. Cross-Sectoral Collaboration: To address the underlying causes of mental health disorders, promote cooperation between the social services, education, employment, and health sectors.

Encourage the development of community-based mental health programs, such as peer support groups, public awareness campaigns, and community education.
Development and implementation of geographically tailored mental health intervention programs should be done in accordance with the high-prevalence areas that have been identified. The goals of these initiatives need to be to increase people's knowledge, accessibility to resources, and choices for treatment in these areas.